\documentclass[preprint]{elsarticle}

\usepackage{siunitx}
\usepackage{multirow}

\usepackage{float}
\usepackage[latin1]{inputenc}   
\usepackage{amsmath} 
\usepackage{epstopdf}           %
\usepackage{flushend}           %
\usepackage{hyperref}           %

\usepackage{amssymb}
\usepackage{lineno}

\usepackage[figuresright]{rotating}
\usepackage{color,soul}

\usepackage{subfigure}

\begin{document}
\begin{frontmatter}

\title{Operando observation of strain relaxation in fatigued pearlitic steel}

\author[First]{Louis Lesage}

\author[Second]{Johan Ahlstr\"{o}m}
\author[Third]{{Yubin Zhang}}
\author[First]{{Can Y{\i}ld{\i}r{\i}m}\corref{cor1}}
\ead{can.yildirim@esrf.fr}

\cortext[cor1]{Corresponding Author}

\address[First]{Experiments Division, European Synchrotron Radiation Facility, 71 Avenue des Martyrs, CS40220, 38043 Grenoble Cedex 9, France.}

\address[Second]{Department of Mechanical Engineering, Division of Computational Mechanics and Materials Engineering, Chalmers University of Technology, Gothenburg SE-412 96, Sweden}
\address[Third]{Department Civil and Mechanical Engineering, Technical University of Denmark, 2800 Kgs. Lyngby, Denmark}

\date{\today}

\begin{abstract}

Pearlitic steels are the material of choice for railway wheels and rails, where their lamellar ferrite-cementite structure balances cost, strength, and wear resistance. However, in use, cyclic loading combined with frictional heat softens the steel, promoting fatigue and eventual failure. Here we use Dark-Field X-ray Microscopy (DFXM) to follow, operando, the grain-scale response of a fatigued pearlitic colony during annealing up to \SI{550}{\celsius}. Orientation maps reveal negligible lattice rotation and no sub-cell formation, indicating that the cementite lamellae suppress long-range dislocation motion. Strain maps nonetheless show pronounced relaxation: the elastic strain spread narrows by nearly \SI{40}{\percent}, with recovery starting below \SI{250}{\celsius}, within the operating temperatures of railway wheels. We attribute this relaxation to short-range annihilation of dislocations confined between lamellae, linking grain-scale strain recovery to the macroscopic softening of pearlitic railway steels in service.

\end{abstract}


\end{frontmatter}

Railway systems are a fast, reliable, and sustainable transport alternative for reducing greenhouse gas emissions. To support this critical infrastructure, materials must meet strict mechanical and durability standards. With increasing train speeds and axle loads, wheel materials face extreme conditions, such as frictional heating from block brakes, which can raise tread temperatures to \SI{550}{\celsius}. Thus, these materials must resist mechanical fatigue and dynamic loads even after high-temperature exposure \cite{walther2004fatigue}.

Low-carbon steels containing around 0.55 wt.\SI{}{\percent}~C are widely used in railway wheels for their cost-effective balance of fatigue and wear resistance \cite{perez1993microstructure, lee2005wear, nikas2016mechanical}.
These properties arise from their microstructure, which consists of pearlite, a lamellar structure of hard cementite and softer ferrite, and pro-eutectoid ferrite grains.
However, prior research has demonstrated that the combination of mechanical fatigue loading followed by thermal cycling can degrade the material's mechanical properties.
This degradation was evidenced by a reduction in hardness, leading to softening, which could eventually lead to fatigue crack formation and potentially catastrophic failure in applications such as railway wheels~\cite{nikas2016mechanical, faccoli2019changes}. 
To develop predictive models for the lifespan and failure risk of railway steels, it is essential to understand the mechanical response of the material under realistic loading conditions. Low-cycle fatigue (LCF) testing provides a well-established method (e.g. ASTM~E606) for examining cyclic plasticity and hardening behaviour. Combined with models of microstructural evolution, it yields data that can be used to predict fatigue damage and residual stresses, including under rolling contact fatigue, which is commonly employed to simulate in-service railway conditions and to analyse the microstructural changes driving mechanical degradation during operation.

Traditional electron microscopy techniques offer very high spatial and angular resolutions; however, they only probe thin volumes just below the surface, resulting in observations further from the actual service conditions of railway steels.
Synchrotron techniques, by contrast, are particularly well-suited for monitoring such phenomena: they enable in situ high-temperature observations, including in bulk samples, and the brightness of the X-ray beam allows for rapid, high-resolution data acquisition.
Yet, prior synchrotron studies have highlighted the challenges of mapping pearlitic materials. Most research has focused on X-ray diffraction and the analysis of diffractograms \cite{taniyama2004structure, ghosh2018deformation}. While these methods provide insights into averaged microstructures, they do not allow the study of local microstructural changes, as 3D imaging techniques could. However, most of these imaging techniques probing orientation and strain levels, such as Laue micro-diffraction \cite{zhang2022stress} or scanning 3DXRD \cite{lesage2026challenges}, struggle to resolve materials with a large intrinsic orientation spread, as is the case for pearlite.

In this study, we employ Dark-Field X-ray Microscopy (DFXM), which addresses these limitations and has demonstrated the ability to image individual crystallographic grains with a spatial resolution down to \SI{150}{\nano\metre}, and to produce orientation maps and strain maps with respective resolutions of \SI{0.001}{\degree} and \num{e-5}~\cite{poulsen2017x}. We monitored, operando, the microstructural evolution within a pearlitic colony from a fatigued railway wheel sample during annealing, the combination of LCF testing, followed by heat treatment, providing a well-controlled yet near-service simulation of the thermomechanical loads experienced by railway steels in service~\cite{ahlstrom2005fatigue, nikas2016mechanical}.

\begin{figure}[ht]
    \centering
    \includegraphics[width=0.65\textwidth]{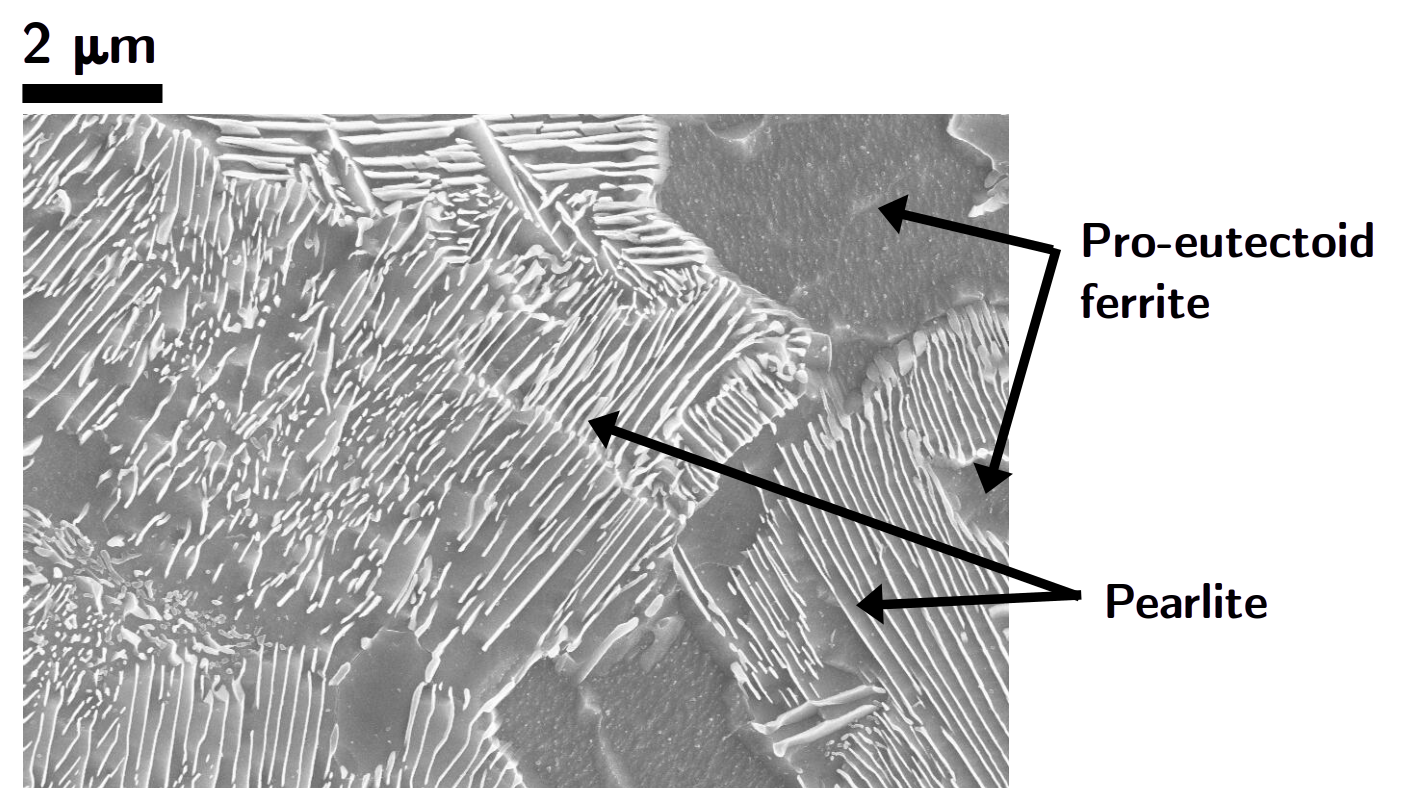}
    \caption{SEM secondary electrons image of a cross-section of a R7T specimen.}
    \label{fig:SEM}
\end{figure}

We investigated the R7T alloy, which is commonly used in the manufacture of railway wheels; its chemical composition is detailed in Table~\ref{tab:composition}. After forging and rolling in production, the wheels are heated to the austenitic range, and rim chilled by spraying water around the periphery of the wheel, yielding a gradient in microstructure and strength and beneficial compressive stresses around the circumference. A slight stress relief is the final part of the industrial process, to even out global residual stress levels within and in between batches.
Due to its composition and heat treatment, the microstructure in the rim parts of the wheel mostly comprises pearlite colonies (alternating ferrite and cementite lamellae with around \SI{125}{\nano\metre} interlamellar spacing) alongside pro-eutectoid ferritic grains as visible on the SEM image in Fig.~\ref{fig:SEM}.

\begin{table}[ht]
\centering
\caption{Chemical composition of the studied pearlitic steel, in wt.\%.}
\label{tab:composition}
\begin{tabular}{lcccccccccc}
\hline
C & Si & Mn & Mo & Cr & Ni & S & P & V & Fe \\
\hline
0.52 & 0.4 & 0.8 & 0.08 & 0.3 & 0.3 & 0.015 & 0.02 & 0.06 & Bal \\
\hline
\end{tabular}
\end{table}

In a previous study, cylindrical specimens, each with a diameter of \SI{6.3}{\milli\metre}, were extracted from the rim of a virgin wheel, with the tested material volume at a depth of approximately \SI{15}{\milli\metre} below the running surface.
The specimens were subjected to push-pull uniaxial LCF tests at room temperature under a constant total strain amplitude of \SI{0.6}{\percent}, resulting in fracture after 4688 cycles. More details on these LCF tests can be found in \cite{ahlstrom2016temperature}.
A needle-shaped sample was subsequently extracted from the gauge section of the fractured specimens using a diamond saw, oriented along the tensile axis. These needles were further thinned via electropolishing to ensure a significant region at the tip had a thickness below \SI{150}{\micro\metre}, ensuring adequate X-ray transmission.

\begin{figure}[ht]
    \centering
    \includegraphics[width=0.5\textwidth]{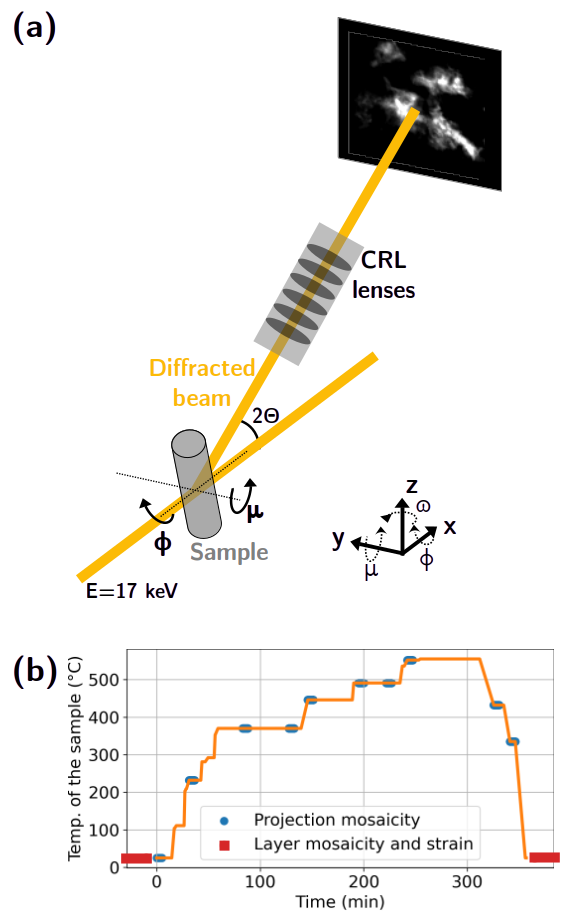}
    \caption{(a) Setup for DFXM at the ESRF ID03 beamline. The near-field detector and the furnace are not shown for simplicity; more information on the heating setup can be found in \cite{lesage2026high}. (b) Temperature profile and timestamps of the scans presented in this study.}
    \label{fig:experimental_setup}
\end{figure}

The DFXM experiments were conducted at the ESRF beamline ID03 \cite{isern2025esrf}. A schematic representation of the DFXM setup is shown in Fig.~\ref{fig:experimental_setup}. The samples were annealed in the beamline-integrated furnace that enables all goniometer rotations and tilts needed for DFXM \cite{lesage2026high}. The furnace set temperature was increased in steps up to \SI{550}{\celsius}. 
The actual temperature of the sample was calculated using the thermal expansion coefficient of ferritic iron \cite{nix1941thermal} and the d-spacing derived from the measured Bragg angle obtained during the diffraction experiments.
DFXM observations of a supposedly single pearlitic colony (for convenience referred to as a grain below) during annealing were performed using a \SI{17}{\kilo\electronvolt} incoming beam, where the (110) reflection of $\alpha$-ferrite occurs at $2\theta = \SI{20.73}{\degree}$.
A near-field PCO.edge detector positioned \SI{107}{\milli\metre} behind the sample was first used to obtain an overview of the diffraction signal and to align the grain to satisfy the diffraction conditions.
Then, the diffracted X-rays were magnified using an objective comprising 87 beryllium compound refractive lenses (CRLs)~\cite{poulsen2017x}. This objective, positioned \SI{273}{\milli\metre} behind the sample, produced a magnified image projected on a far-field PCO.edge detector located \SI{5346}{\milli\metre} downstream of the sample.  
Among the pearlite colonies with the ferrite phase satisfying the diffraction conditions within the tilt range of the DFXM stage and producing a diffracted beam directed toward the far-field detector, the selected colony was the one yielding the highest intensity on the detector to ensure high-quality mapping.

Two types of DFXM images were acquired. (i) Projection imaging, where the quasi-parallel beam was collimated to a box-shaped beam using slits with an opening of \SI{300}{\micro\metre} $\times$ \SI{300}{\micro\metre} illuminating the entire grain and producing a projection image on the detector. (ii) Layer imaging (i.e. magnified section topography), in which the incoming beam was focused into a \SI{500}{\nano\metre}-thick line using a condenser comprising a series of 58 Be lenses. 
Local orientation (mosaicity) variations around the 110 diffraction vector were probed by imaging the grain at multiple tilt positions around the $x$- and $y$-axes (i.e., $\phi$ and $\mu$ angles), as shown in Fig.~\ref{fig:experimental_setup}.a. Axial strain scans along the [110] direction were performed by varying the energy, therefore the wavelength $\lambda$, of the X-ray beam, thereby probing the $d$-spacing and, consequently, the elastic strain within the grains according to Bragg's law. Details on the acquisition parameters for each type of scan can be found in \ref{app:DFXM_parameters}.
The temperature profile applied during annealing and the scans performed are shown in Fig.~\ref{fig:experimental_setup}.b. The applied ramp rate was \SI{50}{\celsius\per\minute}. Layer mosaicity and strain scans were conducted both before and after the temperature cycle, while faster projection mosaicity maps were acquired during the annealing process.
The acquired data were processed using \textit{darfix} software ~\cite{garriga2023darfix}.

\begin{figure}[ht]
    \centering
    \includegraphics[width=0.95\textwidth]{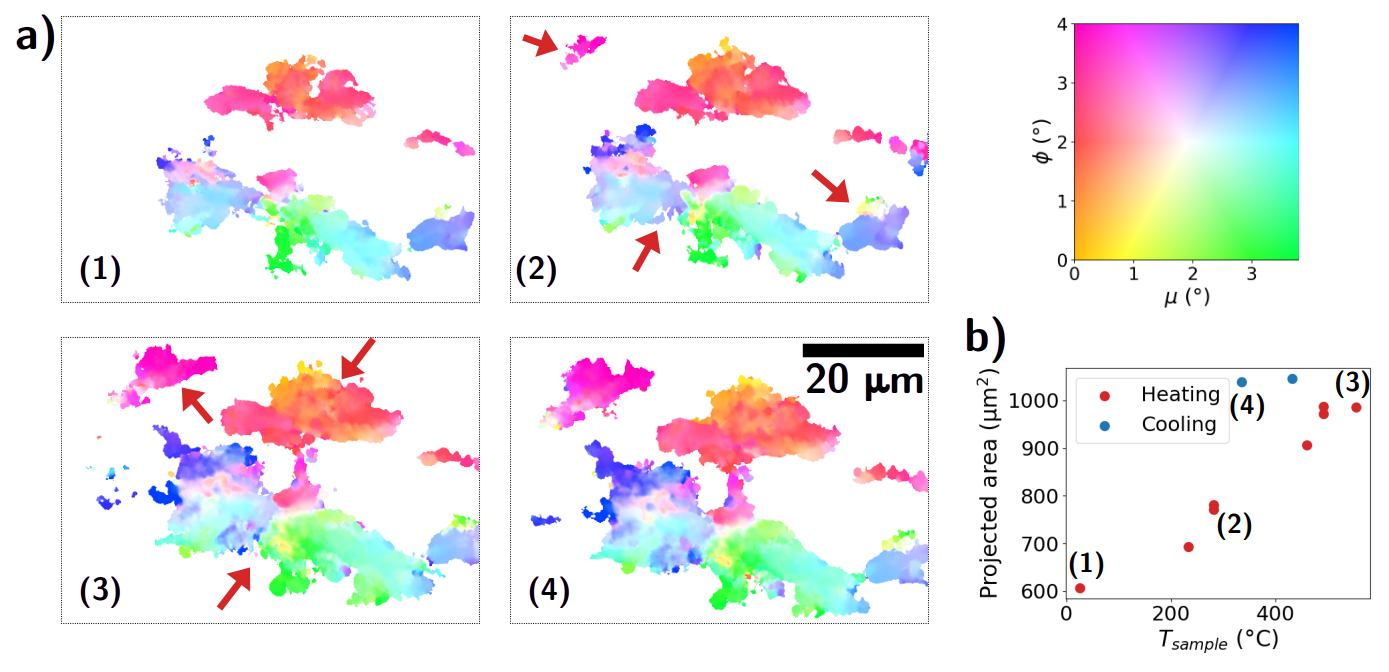}
    \caption{Evolution of the mosaicity maps during annealing/cooling (a) Mosaicity maps of the projected grain at various temperatures. Red arrows indicate new parts of the grain appearing progressively during the temperature cycle. (b) Projected area versus sample temperature.}
    \label{fig:DFXM_proj}
\end{figure}

Orientation maps obtained under identical acquisition conditions for the projected grain at various annealing stages are shown in Fig.~\ref{fig:DFXM_proj}.
The grain exhibits a width exceeding \SI{40}{\micro\metre} and the map bears an orientation spread of \SI{4}{\degree} in both $\mu$ and $\phi$. Both the size and the orientation spread of the grain are larger than what is typically found in pro-eutectoid ferrite~\cite{yildirim20213d}, indicating that we mapped the ferrite phase contained in a pearlitic colony. The cementite lamellae, being a distinct phase, do not contribute to the diffraction signal collected by the aperture of the objective CRLs, rendering them invisible in these maps. Their size is estimated to be around \SI{15}{\nano\metre}, far below the resolution limit of DFXM. Consequently, we did not expect to observe any signal-less stripes corresponding to cementite in the DFXM maps.

The irregular shape of the grain, combined with the broad range of colours in the orientation map, suggests that portions of the grain are absent from these projections. 
However, as illustrated in Fig.~\ref{fig:DFXM_proj}.b, the projected area of the grain on the detector increases progressively during the annealing process up to \SI{550}{\celsius}. This expansion stabilizes during the final heating step and persists upon cooling.
The phenomenon is particularly evident in maps (2) and (3), where new sections of the grain, indicated by red arrows, emerge in the maps. At \SI{280}{\celsius} and \SI{490}{\celsius}, two mosaicity maps were acquired early and late in the temperature plateau (see Fig.~\ref{fig:experimental_setup}.b), with minimal change in the projected area. This suggests that most microstructural changes occur during the temperature ramps.

The progressive filling of the orientation maps upon annealing can be attributed to two non-exclusive hypotheses: (i) a local reorganization of grain orientations, resulting in a narrowing of the total orientation spread within the pearlitic colony: a larger fraction of the grain then satisfies the diffraction condition within the accessible tilt range of the $(\mu, \phi)$ tilt angles, making it possible to map a greater portion of the grain. (ii) the relaxation of elastic strain within the grain, which reduces the distribution of d-spacings across the illuminated volume: more regions of the grain thereby fulfill the Bragg condition such that their diffracted beam falls within the angular acceptance of the objective lens, contributing to the reconstructed image. Projection scans can hardly confirm or refute these hypotheses, as they stack the contributions of a whole grain on a 2D detector. Therefore, layer scans performed before and after annealing were compared to unveil the strain and orientation evolution in various slices of the grain.

\begin{figure}[ht]
    \centering
    \includegraphics[width=0.95\textwidth]{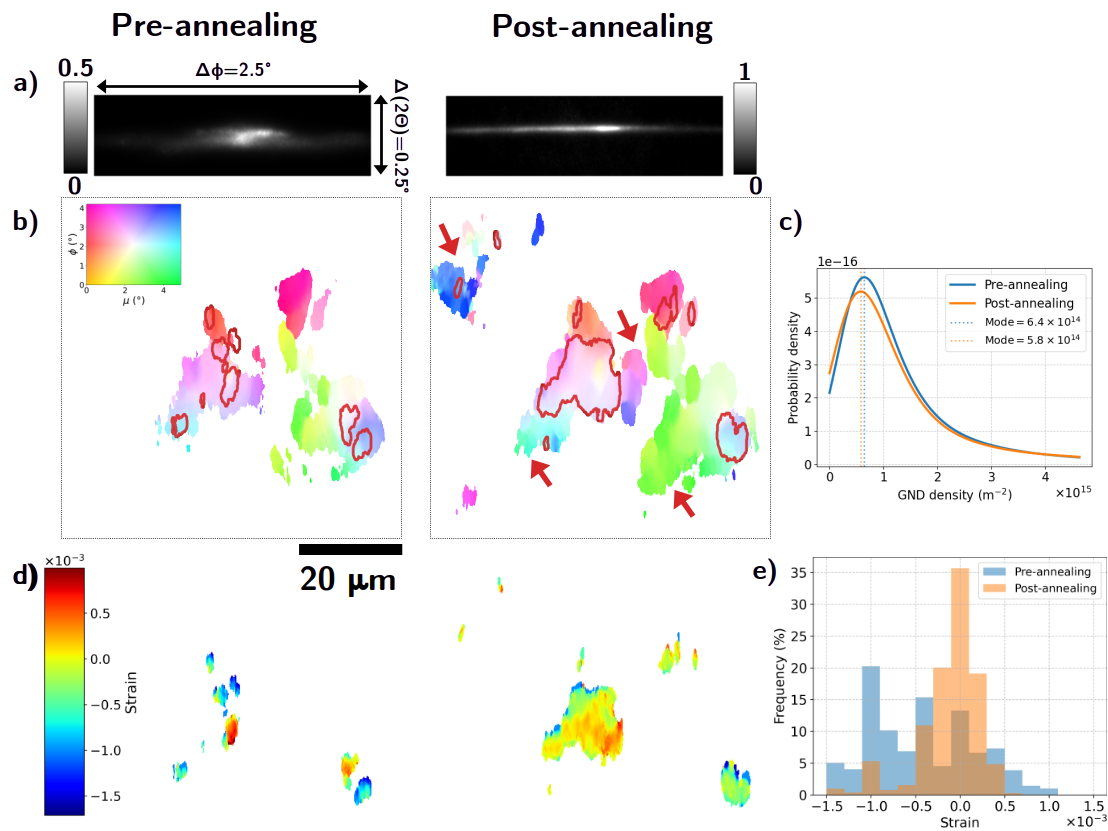}
    \caption{Evolution of grain layers before and after annealing. (a) Portion of the diffraction ring corresponding to the mapped grain, as observed on the near-field camera. (b) Layer mosaicity maps. Red overlays highlight the regions mapped in the strain maps. Arrows show new regions visible after annealing. (c) Probability density of the GND density for all pixels over 10 layers of the grain. The modes of both functions are shown by dotted lines. For each layer, only the portions of the grain present both before and after annealing were considered. (d) Layer strain maps. (e) Frequency of the strain value for all pixels over 10 layers of the grain.}
    \label{fig:DFXM_layers}
\end{figure}

Fig.~\ref{fig:DFXM_layers}.a displays segments of the diffraction ring corresponding to a layer of the mapped grain, integrated over a $\pm \SI{0.5}{\degree}$ range in $\mu$ as observed on the beamline's near-field camera, catching the (110) reflection. 
The observation was done before and after annealing. The width of the detector covers a $\phi$ range of \SI{2.5}{\degree}. A notable reduction in the ring's thickness is evident, with the full width at half maximum (FWHM) decreasing from \SI{0.033}{\degree} to \SI{0.011}{\degree}. These values can hardly be converted to strain, as the FWHM convolutes both strain and grain thickness. 
The grain shape and size are expected to remain unchanged at the studied temperatures, so their contribution to the FWHM is constant. This supports the hypothesis of a reduction in elastic strain during annealing. The increased relative intensity of the ring post-annealing further supports a reduction in the $2\theta$ spread, as the intensity is concentrated over a smaller $2\theta$ range.

To determine whether any concurrent orientation changes occurred, we conducted orientation map scans across multiple layers of the same grain, before and after annealing. Fig.~\ref{fig:DFXM_layers}.b shows orientation maps for one of these layers located in the center of the scanned area. In agreement with the previous projection scans (Fig.~\ref{fig:DFXM_proj}), the mapped area is larger post-annealing, with the appearance of new features as indicated by red arrows. The zones that are recognizable in both figures have similar colors, suggesting that they did not undergo a visible orientation change during the annealing process. In order to track local changes within the grain, we calculated for each pixel the local orientation difference $\Delta\theta = \sqrt{\Delta\mu^2 + \Delta\phi^2}$~\cite{ahl2017ultra, cretton2025observation}, where $\Delta\mu$ and $\Delta\phi$ represent the average orientation difference with the eight directly neighboring pixels. 
This orientation difference is proportional to the geometrically necessary dislocation (GND) density \cite{wilkinson2010determination, moussa2015quantitative, yildirim20253d}, which can be calculated as:
\begin{equation}
    \rho_{\text{GND}} = \frac{\alpha \Delta\theta}{b x} \, \cite{moussa2015quantitative},
\end{equation}
where $\alpha = 2$ for tilt misorientation, $b = \SI{0.247}{\nano\metre}$ is the magnitude of the Burgers vector for BCC iron, and $x = \SI{175}{\nano\metre}$ is the distance to neighbouring pixels.
Fig.~\ref{fig:DFXM_layers}.c shows the frequency of the GND density for all pixels over 10 layers before and after annealing. To ensure a strict comparison before and after annealing, only the portions of the grain present in both conditions were taken into account for this calculation. We observe that, after annealing, the distribution tends to slightly shift towards smaller GND density values, with a reduction of 10~\% in the mode of the probability density function, suggesting that annealing induced a local decrease in misorientation and GND density.

Fig.~\ref{fig:DFXM_layers}.d presents strain maps for the same layers as those shown in the mosaicity maps. The mapped area is significantly smaller, as only the $\mu$ angle was integrated during these scans due to beam-time constraints. Consequently, regions of the grain that diffract at $\phi$ angles deviating from the central value were omitted. The central $\phi$ value in the strain scans was adjusted after annealing to align with the newly visible regions in the mosaicity maps following heat treatment. This adjustment accounts for the observed differences in the regions mapped in the strain scans before and after annealing. Importantly, these differences do not arise from changes in the absolute $\phi$ values of the regions, as the mosaicity maps confirm their stability.
Fig.~\ref{fig:DFXM_layers}.e presents the strain distribution for all pixels across ten layers, scanned both below and above the layer shown in the other panels, over a distance of \SI{50}{\micro\metre}. The strain values were calculated relative to the most frequent $d$-spacing observed in the sample post-annealing, which was treated as the strain-free reference, given the expected strain relaxation during annealing.
Prior to annealing, the strain distribution appears significantly broader, with the pre-annealing state exhibiting predominantly compressive strain (relative to the post-annealing reference) following the LCF test.
The post-annealing distribution demonstrates a notable reduction in the elastic strain spread, with the standard deviation decreasing by almost 40~\%. Both compressive and tensile strains appear to relax, as the distribution recenters around the strain-free value.

The relatively high strain levels identified in the grain of interest prior to annealing can be attributed to lattice defects introduced either during pearlite formation or during the fatigue test. In the latter case, the expected increase in dislocation density at this strain level is primarily due to threading dislocations bulging into the ferrite lamellae~\cite{nikas2018evaluation,dhar2020multi}.
These defects cause lattice distortion, which manifests as broadening of the diffraction peaks, as observed in Fig.~\ref{fig:DFXM_layers}.a, and a broader distribution of elastic strain throughout the observed grain, as in Fig.~\ref{fig:DFXM_layers}.e. The observed strain in the grain of interest appears to be predominantly compressive. However, there is no evidence suggesting that this observation can be generalized to all grains in the sample. Reproducing the study across several grains may help clarify this point, yet it remains beyond the scope of this work.

The GND probability density function in Fig.~\ref{fig:DFXM_layers}.c shows little reduction in GND density following annealing. At longer ranges, no significant change in lattice orientation is observed in Fig.~\ref{fig:DFXM_layers}.b. Furthermore, no sub-cell formation is detected, as might be expected during a recovery process \cite{humphreys1995recrystallization}. Such a process may not be able to occur, as previous studies have shown that dislocation motion can be impeded over large distances, with the lamellae of the structure acting as barriers to dislocations \cite{li2003insitu,ghaffarian2016nanoindentation}. 
This suggests that the dislocation activity is confined within the interlamellar spacing of ferrite, typically below the DFXM resolution of \SI{150}{\nano\metre}, hence, any sub-cell formation would remain undetectable.
These limited GND activity and orientation changes strongly indicate that the increase in projected area observed in Fig.~\ref{fig:DFXM_proj} arises from a reduction in elastic strain rather than lattice rotations in the grain of interest. This reduction in elastic strain is clearly seen in Fig.~\ref{fig:DFXM_layers}.a and .e for the post-annealing sample, where the FWHM of the strain distribution has narrowed, indicating recovery.

\begin{figure}[ht]
    \centering
    \includegraphics[width=0.5\textwidth]{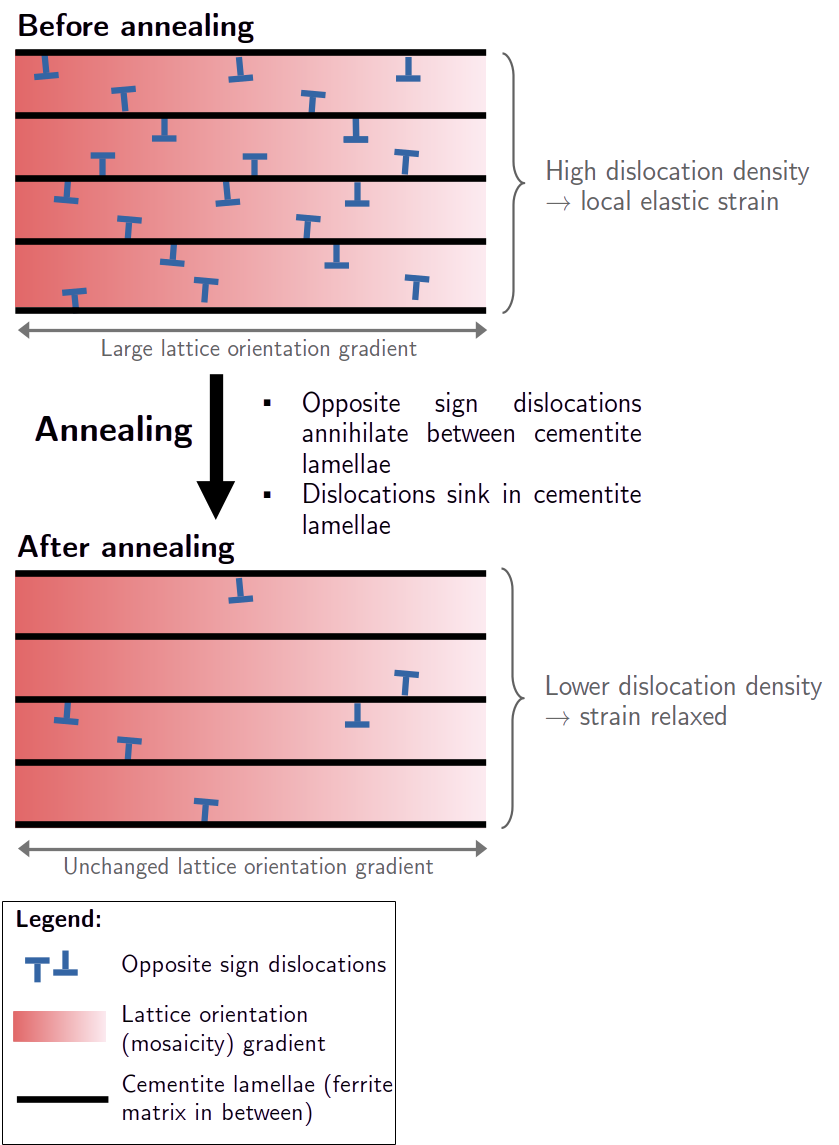}
    \caption{Proposed mechanism upon annealing: strain relaxation without long-range lattice rotation.}
    \label{fig:mechanism}
\end{figure}

The narrowing of the FWHM in the strain distribution can therefore be primarily attributed to a reduction in dislocation density resulting from the annihilation of dislocations with opposite signs between cementite lamellae, or their migration to interfaces. Notably, ferrite-cementite interfaces are known to act as dislocation sinks, as demonstrated in prior works \cite{chen2016defect,janakiram2021new,liang2020dislocation}. Such dislocation activity is facilitated by the substantial reduction in the energy barrier for dislocation glide and climb at elevated temperatures.
Fig.~\ref{fig:mechanism} summarizes the proposed mechanisms underlying the observed strain relaxation. Interestingly, this strain reduction was not detectable in the Debye-Scherrer rings of all grains, as captured before and after annealing using an area detector (\ref{app:Debye-Scherrer}). This limitation highlights the resolution constraints of the technique and underscores the necessity of higher-resolution methods, such as DFXM.

Finally, we reveal that the microstructural changes driven by annealing remain steady throughout the temperature ramp, as evidenced in Fig.~\ref{fig:DFXM_proj}.b. Above \SI{500}{\celsius}, spheroidisation of pearlite, whereby cementite lamellae transform into globular structures, is expected, leading to a further decrease in hardness~\cite{cvetkovski2011thermal}. Given the temperature ramp of our heat treatment, some spheroidisation likely occurred towards the end of the thermal loading. Whilst the layer maps acquired pre- and post-annealing do not allow the contributions of dislocation density decrease and spheroidisation to be decoupled, our observations confirm that strain reduction begins even before any spheroidisation event, at temperatures as low as \SI{250}{\celsius}.
This aligns with previous works highlighting a decrease in hardness in pearlitic railway steels with increasing annealing temperature~\cite{nikas2016mechanical}, indicating a correlation of the observed elastic strain decrease with the macroscopic hardness reduction.

 In summary, our study provides grain-scale observations of the strain relaxation in pearlitic steel under thermal loading following fatigue testing. Using DFXM, we demonstrate a steady decrease in elastic strain throughout the temperature ramp. Such grain-scale relaxation might otherwise be averaged out or remain undetected by traditional diffraction techniques, underscoring the potential of DFXM.

The absence of sub-cell formation during annealing suggests that dislocation mobility in pearlite is restricted to the interlamellar spacing. The decrease in elastic strain is primarily attributed to a reduction in dislocation density constrained between cementite lamellae. These microstructural changes initiate below \SI{250}{\celsius}, a temperature range routinely reached by railway wheels in operation. These grain-scale insights, both qualitative and quantitative, pave the way for predictive models to enhance the lifecycle and safety of pearlitic steels in industrial applications.

\section*{Acknowledgements}

The authors thank the ESRF for the provision of beamtime at ID03 under the proposal No. IHMA-752. 

\section*{Disclosure statement}
The authors report there are no competing interests to declare.

\section*{Funding}

CY and LL acknowledge the financial support from the ERC Starting Grant No. 10116911.

\bibliography{references}


\appendix
\section{DFXM acquisition details}
The table below shows the scanning parameters used during DFXM acquisitions.
\label{app:DFXM_parameters}

\begin{table}[h!]
\centering
\caption{DFXM scanning parameters.}
\label{tab:DFXM_params}
\footnotesize
\setlength{\tabcolsep}{4pt}
\begin{tabular}{lccccc}
\hline
\textbf{Scan type} & \textbf{Fig.} & \shortstack{\textbf{$\mu$ step} \\ \textbf{(\si{\degree})}} & \shortstack{\textbf{$\chi$ step} \\ \textbf{(\si{\degree})}} & \shortstack{\textbf{Energy step} \\ \textbf{(keV)}} & \shortstack{\textbf{Exposure} \\ \textbf{time (s)}} \\
\hline
Projection mosaicity & \ref{fig:DFXM_proj} & 0.2 & 0.333 & - & 1 \\
Layer mosaicity & \ref{fig:DFXM_layers}b & 0.1 & 0.1 & - & 0.3 \\
Layer strain & \ref{fig:DFXM_layers}d & 0.1 & - & $8.3 \times 10^{-3}$ & 0.3 \\
\hline
\end{tabular}
\end{table}

\section{Debye-Scherrer rings acquisition}
\label{app:Debye-Scherrer}
\subsection{Data acquisition and analysis}
Debye-Scherrer rings were acquired using a FReLoN CCD detector with a resolution of 2048\(\times\)2048 pixels and a pixel size of \SI{47.3}{\micro\meter}, positioned \SI{283}{\milli\meter} downstream of the sample. The beam energy was set to \SI{55}{\kilo\electronvolt} to capture two ferrite diffraction peaks, and its size was \SI{300}{\micro\metre} $\times$ \SI{300}{\micro\metre} to probe a significant volume of the sample.
For more statistics, Debye-Scherrer rings were acquired each time for each \SI{1}{\degree} step of the \SI{360}{\degree} possible $\omega$ rotation. These rings were then azimuthally integrated over \SI{360}{\degree} using the Python library \texttt{pyFAI}~\cite{kieffer2020new, ashiotis2015fast} to get 1D diffractograms of intensity versus scattering angle.

\begin{figure*}[ht]
    \centering
    \includegraphics[width=0.5\textwidth]{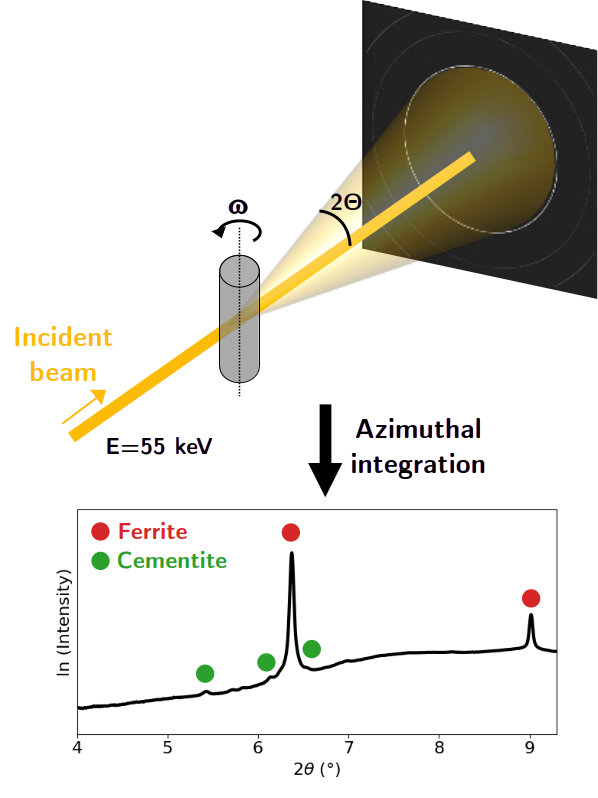}
    \caption{Setup for the Debye-Scherrer ring acquisition at the ESRF ID03 beamline. The presented diffractogram was acquired as a pre-characterisation of the sample before annealing.}
    \label{supfig:experimental_setup_DebyeScherrer}
\end{figure*}

The azimuthal integration of the initial Debye-Scherrer rings attests to the coexistence of the ferritic phase with cementite. The setup and the integrated diffractogram are shown in Fig.~\ref{supfig:experimental_setup_DebyeScherrer}.

\subsection{Operando monitoring of Debye-Scherrer rings}
We applied the same temperature ramp to a second sample as used for the DFXM-analysed sample and followed the evolution of the Debye-Scherrer rings operando.

\begin{figure*}[ht]
    \centering
    \includegraphics[width=0.5\textwidth]{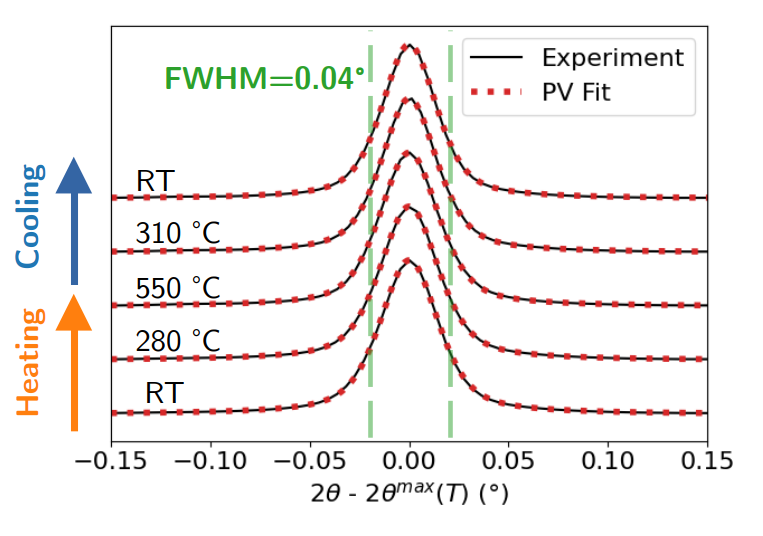}
    \caption{Azimuthal integration of the ring over \SI{360}{\degree} rotation of the sample for the (110) peak. The peak position was centered on 0 to remove any contribution of the thermal expansion.}
    \label{supfig:XRD_patterns}
\end{figure*}

Fig.~\ref{supfig:XRD_patterns} shows the (110) peak in the diffractograms, which were integrated over the diffraction rings and summed for a full 360$^\circ$ rotation around $\omega$. A Pseudo-Voigt fit of the peak is overlaid on the integrated data. All peaks were centred at 0 to correct for thermal expansion-induced shifts, and their intensities were normalized. After these corrections, all peaks exhibit identical symmetry and width (FWHM = \SI{0.04}{\degree}). This FWHM value matches that obtained from integrating the Debye-Scherrer ring of the Si calibrant, considered a strain-free reference. Additionally, the pixel size of the FReLoN CCD detector limits the strain resolution of the technique to a strain level of 0.001, similar to the maximum strain probed by DFXM in our grain of interest. Smaller variations in the strain levels are therefore undetectable with this camera, justifying the use of higher strain-resolution techniques such as DFXM.

\end{document}